\documentclass[a4paper,preprint]{aastex}

\def\spose#1{\hbox to 0pt{#1\hss}}
\def\lta{\mathrel{\spose{\lower 3pt\hbox{$\mathchar"218$}}
     \raise 2.0pt\hbox{$\mathchar"13C$}}}
\def\gta{\mathrel{\spose{\lower 3pt\hbox{$\mathchar"218$}}
     \raise 2.0pt\hbox{$\mathchar"13E$}}}

\begin{document}

\title{THE ENERGY OF LONG DURATION GRBS}

\author{T. Piran\footnote{tsvi@phys.huji.ac.il}}
\affil{Racah Institute, The Hebrew University, Jerusalem 91904,
Israel}
\author{P. Kumar\footnote{pk@ias.edu}}
\affil{IAS, Princeton, NJ 08540, USA}
\author{A. Panaitescu\footnote{alin@astro.princeton.edu}}
\affil{Dept. of Astrophysical Sciences, Princeton University, NJ
08544, USA}
\author{L. Piro\footnote{piro@ias.rm.cnr.it}}
\affil{Istituto Atrofisica Spaziale, CNR, Via Fosso del Cavaliere,
Rome 00133 Italy}

\begin{abstract}
The energy release in gamma-ray bursts is one of the most useful
clues on
 the nature of their ``inner engines". We show that, within the framework
 of the relativistic external shocks afterglow model, the narrowness of the observed
 X-ray luminosity of GRB afterglows, implies that the energy of the GRB jets
 after the early afterglow phase spans less than one order of magnitude.
 This result is not affected by uncertainties in the electrons energy,
 magnetic field strength, and external medium density. We argue that the
 afterglow kinetic energy is within a factor of two of the initial energy
 in the relativistic ejecta, therefore the energy output of the central
 engine of long duration GRB has an universal value.
\end{abstract}

Subject headings: gamma-rays: bursts - ISM: jets and outflows -
radiation mechanisms: non-thermal - shock waves


In the last three years we have learned a great deal about long
duration gamma-ray bursts (GRBs). The Italian-Dutch satellite
BeppoSAX provided angular position of several dozen long bursts
to within about 3 arc-minutes which enabled follow up
observations  in the x-ray (see Piro, 2000), optical, milli-meter
and radio frequencies which has provided a wealth of information
on these explosions. These observations are described well by the
relativistic fireball model (see e.g. Piran, 1999). According to
this model the energy from the central source is deposited in
material that moves with speed very close to the speed of light.
The kinetic energy of this material is converted to the observed
electromagnetic radiation as a result of collisions between fast
moving material that catches up with slower moving ejecta, and
the shock heated circum-burst medium. The nature of the ``inner
engines" that expels relativistic material, which is responsible
for the the GRBs, is not determined yet. The energy of the
relativistic matter ejected by the ``inner engine", $E_{rel}$, is
one of the most important clues on its nature. Our goal is to
find a reliable estimate of $E_{rel}$.

Given an observed $\gamma$-ray fluence and the redshift to a
burst one can easily estimate the energy emitted in $\gamma$-rays,
$E_{\gamma,iso}$  assuming that the emission is isotropic.
$E_{\gamma,iso}$ can also be estimated from the BATSE catalogue
by fitting the flux distribution to theoretical models (Cohen \&
Piran, 1995; Schmidt, 2001). As afterglow observations proceeded,
alarmingly large values (Kulkarni et al. 1999) ($3.4 \times
10^{54}$ergs for GRB990123) were measured for $E_{\gamma,iso}$.
However, it turned out (Rhoads, 1999; Sari Piran \& Halpern,
1999) that GRBs are likely beamed and $E_{\gamma,iso}$ would not
then be a good estimate for the total energy emitted in
$\gamma$-rays. We define instead: $E_\gamma =
(\theta^2/2)E_{\gamma,iso}$. Here $\theta$ is the effective angle
of $\gamma$-ray emission, which can be estimated from $t_{b}$,
the time of the break that appears latter in the afterglow light
curve (Rhoads, 1999; Sari Piran \& Halpern, 1999): $\theta =0.12
(n/E_{51})^{1/8} t_{b,days}^{3/8}$,  where $E_{51}$ is the
isotropic-equivalent energy kinetic energy during the adiabatic
fireball phase, discussed below, in units of $10^{51}$ergs, and
$t_{b,days}$ is the break time in days. Recently Frail et al.
(2001, hereafter F1) estimated $E_\gamma$ for 18 bursts, finding
typical values around $10^{51}$ergs. While $E_\gamma$ is closer to
$E_{rel}$ it is still not a good estimate.  First, we have to
take an unknown conversion efficiency of energy to $\gamma$-rays
into consideration: $E_{rel} = \epsilon^{-1} E_\gamma
=\epsilon^{-1} (\theta^2/2) E_{\gamma,iso}$. Second, the large
Lorentz factor during the $\gamma$-ray emission phase, makes the
observed $E_\gamma$ rather sensitive to angular inhomogeneities
of the relativistic ejecta (Kumar \& Piran, 2000).

We consider here another quantity: $E_{K,ad}$, the kinetic energy
of the ejecta during the adiabatic afterglow
phase\footnote{\baselineskip 9pt The external relativistic shock
becomes adiabatic about 1/2 an hour after the explosion, and
furthermore the loss of energy during the earlier radiative phase
is typically not large.}. Clearly: $E_{rel} \ge
\overline{E_\gamma} + E_{K,ad}= \epsilon E_{rel} + E_{K,ad}$,
where $\overline{E_\gamma}$, is the angular average of
$E_\gamma$. The inequality arises from possible energy losses
during the early afterglow radiative phase. However observations
of long time tails of GRBs suggest that, unless this energy is
radiated at extremely high energy channel, this losses are not
large (Burenin et al, 1999, Giblin et al, 1999; Tkachenko et al.,
2000) . Therefore, with $\epsilon \approx 10\%$ (Kobayashi, Piran
\& Sari, 1997) ($\epsilon$ cannot be too close to unity otherwise
there won't be afterglow) we expect that $E_{K,ad} \approx
E_{rel}$ to within better than a factor of 2. Hereafter we drop
the subscript $kin$ denoting $E=E_{K,ad}$.

The purpose of this paper is to determine the spread of $E$ using
the x-ray afterglow flux. The advantage of the method presented
here is that it is independent of the uncertain density of the
ISM, and in fact all other parameters, except for the observed
width of the distribution of the jet opening angle.

One way of determining how $E$ is distributed is to figure out the
energy for individual bursts by modeling their afterglow emission
over a wide range of frequency and time. This procedure, carried
out for 8 well studied bursts (Panaitescu \& Kumar 2001,
hereafter PK01), gives the mean energy to be $\sim
3\times10^{50}$ergs and the standard deviation of the log of
energy distribution to be about 0.3. However, the detailed
modeling of individual GRB afterglow emission is cumbersome and
time consuming, and hard to carry out for a large sample of
bursts especially considering that we need data in radio, optical
and x-ray bands with good time coverage for this kind of an
analysis to be useful.

Moreover, this procedure is not necessary if we simply want to
know the width of energy distribution. For this we can use the
x-ray afterglow flux at a fixed time after the explosion. The
width of the distribution of this flux, an easily measurable
quantity, yields the width of the distribution for the energy
release in the explosion. This method is described below. The
observation of the x-ray flux should be carried out at a
sufficiently late time such that the angular variations and
fluctuations across the surface of the ejecta have been smoothed
out. This occurs several hours after the explosion when the bulk
Lorentz factor of the ejecta has decreased to about 10 at which
time we see a good fraction of the relativistic ejecta.
Conveniently, this is also when the observed x-ray in the 2-10
keV band is above the cooling frequency in which case the
observed flux is independent of the density of the medium in the
vicinity of the explosion (Kumar 2000). Furthermore, it is best to
carry out observations while $\Gamma > \theta^{-1}$, that is
before the jet begins its sideways expansion, which makes the
interpretation of the observed flux much simpler. In five cases
of GRB afterglow light curves where we see the effect of the
finite opening angle of explosion as steeping of the light curve,
we find the effect manifests itself at least one day after the
explosion. Thus, the above two requirements suggest it is best to
consider the x-ray afterglow flux between several hours and a day
after the explosion.

The x-ray afterglow fluxes from GRBs have a power law dependence
on $\nu$ and on the observed time $t$ (Piro, 2000): $f_\nu(t)
\propto \nu^{-\beta} t^{-\alpha}$ with $\alpha \sim 1.4$ and
$\beta \sim 0.9$. The observed x-ray flux per unit frequency,
$f_x$, is related, therefore, to, $L_x$, the isotropic luminosity
of the source at redshift, z by:
\begin{equation}
L_x(t) = {4 \pi d_L^2 \over (1+z)} f_x(t)(1+z)^{\beta-\alpha}
\equiv f_{x}(t) Z(z) \ ,
\end{equation}
where $Z(z)$ is a weakly varying function of $z$. For bursts with
$0.5<z<4$ and with $\beta-\alpha \approx -0.5$ we find $\sigma_Z
\approx 0.31$ (for a cosmology with $\Omega_m=0.3$ and
$\Omega_\Lambda=0.7$). Here and thereafter we denote by
$\sigma_X$ the standard deviation of the $\log(X)$, unless noted
otherwise.

Assuming that the x-ray luminosity does not evolve with redshift
we can relate the dispersion of $\log(L_x)$ at a fixed observer
time after the explosion, $\sigma_{L_x}$, with the observed
dispersion $\sigma_{f_{x}}$: $\sigma^2_{L_x}=\sigma^2_{f_{x}} +
\sigma^2_Z\approx \sigma^2_{f_{x}} \ .$ Using 21 BeppoSAX bursts
(Piro, 2000) we find $\sigma_{f_{x}}\approx 0.43 \pm 0.1$ (see the
caption for Figs. 1 and 2 for the details of the observations and
the analysis), and therefore, $\sigma_{L_{x}} \approx 0.43$ to
within 25\%. This result is supported by 10 x-ray light-curves of
GRBs with known red-shifts, and $\alpha$ and $\beta$.

The x-ray flux, in an energy band above the cooling frequency at a
fixed time after the burst, depends on the energy per unit solid
angle in the explosion (provided that $\Gamma> \theta^{-1}$ at the
time of observation), and on the fractional energy taken up by
electrons, $\epsilon_e$. The flux does not depend on the density
of the surrounding medium, $n$, or its stratification or the
fractional energy in the magnetic field, $\epsilon_B$ (Kumar,
2000). The flux has a weak dependence on $n$ when the electron
cooling is dominated by the inverse Compton scattering
(Panaitescu \& Kumar 2000).

Under these, rather general, conditions the standard synchrotron
fireball model implies that the isotropic equivalent flux at
frequency $\nu$ above the cooling frequency, at a fixed elapsed
time since the explosion, is given by (Kumar 2000; Freedman \&
Waxman, 2001):
\begin{equation}
L_x = \eta_p \left[{d E\over d\Omega}\right]^{(p+2)/4}
   \epsilon_e^{p-1}\epsilon_B^{(p-2)/4},
   \label{LX}
\end{equation}
where $dE/d\Omega$ is the energy per unit solid angle, and
$\eta_p$ is a constant\footnote{\baselineskip=10pt It is worth
emphasizing that equation (2) is independent of the details of
the jet structure i.e., the variation of Lorentz factor across
the jet, since at about 1/2 a day after the explosion the jet
Lorentz factor has dropped to $\sim10$ and we see a good fraction
of the entire jet surface.}. Assuming that there is no correlation
between the microscopic variables, $\epsilon_e$, $\epsilon_B$,
$p$ and $dE /d\Omega$ we obtain from the above equation that
$\sigma_{dE/d\Omega}<\sigma_{L_x}$. Using $\sigma_{L_x}\approx
0.43\pm 0.1$ for the 21 BeppoSAX bursts we find that
$\sigma_{dE /d\Omega}\leq 0.43\pm 0.1$.

 From $\sigma_{dE /d\Omega}$ we can now obtain $\sigma_E$ that
characterizes the distribution of the kinetic energy provided we

know $\sigma_\theta$ using the trivial relation: $\sigma^2_{dE
/d\Omega} = \sigma^2_{E}+4 \sigma^2_\theta$. Panaitescu \& Kumar
(PK01) and Frail et al. (F01) have estimated the jet opening
angles from the observed (or lack of) breaks in the light curves
of optical afterglow light curves. For 8 GRBs from the PK01
sample we have: $\sigma_\theta \approx 0.31\pm 0.06$ while a
sample of 10 bursts from F01 yields $\sigma_\theta \approx 0.28\pm
0.05$. If these values are representative for the whole GRB
population we find a marginally viable solution within two
$\sigma$ errors of $\sigma_E < 0.2$ (for the PK01  result) and
$\sigma_E < 0.27$ (for the F01 data); to get a viable solution we
had to take both the values of $\sigma_{L_x}$ one standard
deviation above the mean and the value of $\sigma_\theta$ one SD
below the mean. This result suggests that there is a narrow
energy distribution; the FWHM of $E$ being less than a factor of
5. If $E$ and $\theta$ are correlated the above relation is
modified i.e. $\sigma^2_{dE/d\Omega} = \sigma^2_{E}+
4\sigma^2_\theta -4\zeta\sigma_E\sigma_\theta$. Both the PK01 and
the F01 data  show that this correlation ($\zeta$) if non-zero is
weak, less than 0.35. With such a correlation the allowed  energy
distribution could be somewhat (but not significantly) broader.

A stronger constrain on $\sigma_E$ can be obtained using
$\sigma_{\epsilon_e}=0.3$ for 8 GRBs analyzed by PK01. It follows
from Eq. \ref{LX} that a non-zero value for $\sigma_{\epsilon_e}$
makes the distribution for $dE/d\Omega$ and hence $E$ even
narrower, however to quantify this effect we need a larger data
set.

We have argued before that $E=E_{K,ad}$, discussed here, is a
rather good estimate to $E_{rel}$ the total energy emitted by the
``inner engine". The constancy of $E_{K,ad}$ is another indication
for it being a good measure of $E_{rel}$. The constancy of
$E_{K,ad}$ is also an indication that the assumptions that have
lead to Eq. \ref{LX} are justified. Otherwise it would have been
remarkable if starting from different levels of initial energy
and having different amounts of energy losses the final kinetic
energy of the afterglow would converge to a constant value. At
present there is no way to tell whether significant amounts of
additional energy is released in other forms, e.g. non
relativistic particles or neutrinos.  Similar arguments suggest
further that this total energy emitted by the ``central engine"
does not vary significantly and that it is rather close to the
energy estimated here.

The distribution of relativistic kinetic energy during the
afterglow phase is narrow, with full width at half maximum less
than one decade. These results suggest that the wide distribution
of directly and indirectly determined $E_{\gamma,iso}$ results
from the distribution of beaming angles, from a variation of
$dE/d\Omega$ across the jet, and from a variable efficiency in
conversion of kinetic energy to $\gamma$-rays. The fact that GRB
engines are "standard" engines in terms of their energy output
provide a very severe constraint on the nature of these enigmatic
explosions. For instance, in the collapsar model for GRBs the
central engine is composed of a black hole (BH) and an accretion
disk around it (Woosley 1993; Paczynski, 1998; MacFadyen \&
Woosley, 1999). This model has two energy reservoirs which can be
tapped to launch a relativistic jet: the BH rotation energy and
the gravitaional energy of the disk. Our result of nearly
constant energy in GRBs implies that the mass accretion on to the
BH plus the possible conversion of rotational energy of the BH to
kinetic energy of the jet does not vary much from one burst to
another inspite of the fact that both the disk mass and the BH
spin are expected to vary widely in the collapse of massive stars.

BeppoSAX is a mission of the Italian Space Agency (ASI) with
participation of the Dutch space agency (NIVR). TP acknowledges
support by the US-Israel BSF.

\clearpage

\begin{figure}
\includegraphics{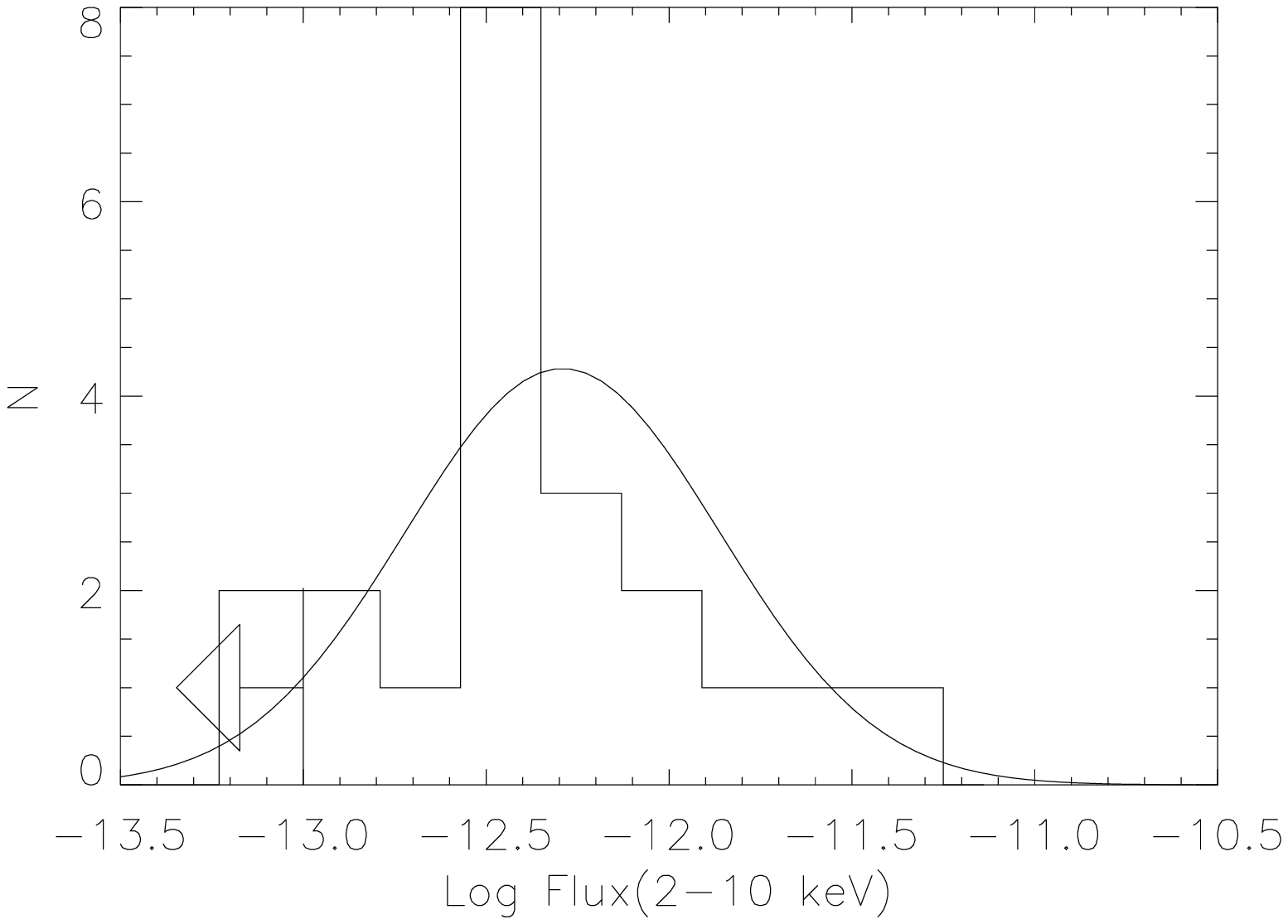}
\caption[]{\footnotesize{The distribution of X-ray fluxes (2-10
keV) at t=11 hours after the GRB in  21 afterglows observed by
BeppoSAX. The sample includes all the fast observations performed
by BeppoSAX on GRB from January 1997 to October 1999. Data are
from Piro (2001, proc. of X-ray astronomy 99, and references
therein) Stratta et al. (2001, in preparation), Nicastro et al.
1999 (IAUC7213), Feroci et al. (2001 A\&A, in press), Montanari
et al. (2001, Proc. of the 2nd wokshop on GRB in the afterglow
era, Rome, 17-20 Oct. 200, E. Costa, F. Frontera J. Hjort eds.,
in press), Piro et al. (1999, GCN 409) and \'T Zand et al. (2000,
ApJ, 545, 266). No X-ray afterglow was detected in GB990217 (Piro
et al. 1999, IAUC 7111) to the limiting instrumental sensitivity
of $10^{-13}$ erg cm$^{-2}$ s$^{-1}$, 6 hours after the burst. In
the case of GB970111 (Feroci et al. 1998 A\&A 332,L29) a
candidate was detected, but evidence of fading behaviour is
marginal, so we have considered both cases as upper limits
(indicated by the arrow in fig).}}
\end{figure}
\begin{figure}
\includegraphics{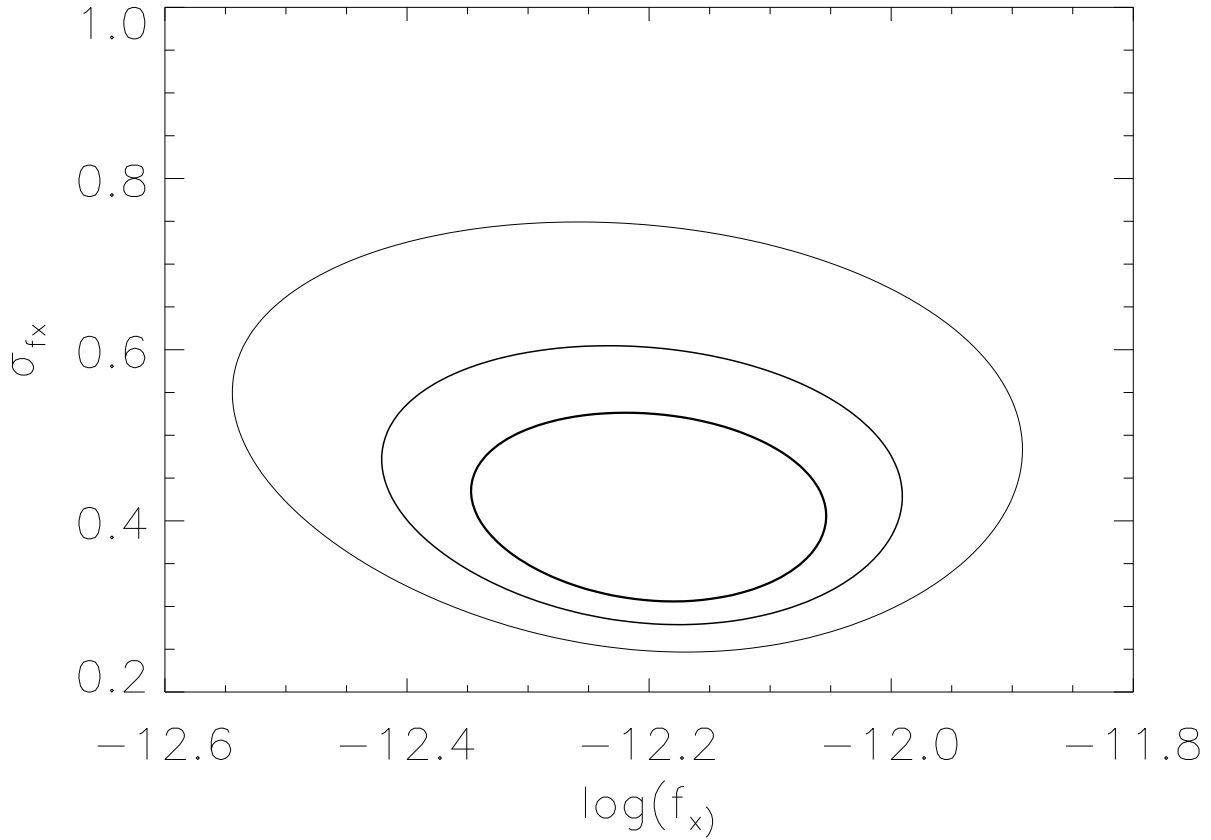} \caption[]{\footnotesize{Likelihood contour lines
(corresponding to 99\%, 90\% and 69\%  confidence levels) in the
$\overline{\log(f_x)}$, $\sigma_{f_x}$ plane for the X-ray flux
distribution as inferred from 21 GRBs detected by BeppoSAX from
January 1997 to October 1999. We determine $\overline{\log(f_x)}$
and $\sigma_{f_x}$ by minimizing the likelihood function
$S=-\sum_i \ln\left\{\left[2 \pi(\sigma_i^2 +
\sigma^2_{f_x})\right]^{-1/2} \exp\left[ - (\log f_{x_i} -
\overline{\log f_x})^2/2 (\sigma_i^2 +
\sigma^2_{f_x})\right]\right\}$; where $\log f_{x_i}$ and
$\sigma_i$ are the observed x-ray flux 11 hr after the onset of
the i-th GRB and the associated measurement error respectively.
The maximal likelihood is at $\overline{\log(f_x)}=-12.2{\pm0.2}$
and $\sigma_{f_x} =0.43^{+.12}_{-.11}$. Two upper limits of
$10^{-13}$ ergs cm$^{-2}$ s$^{-1}$ in the 2--10 kev band at 11
hours after the bursts are included in this data set. The value
of $\sigma_{f_x}$ is 0.42 if we exclude these upper limits.  We
have checked a posteriori with a Kolgomorov-Smirnov test that the
distribution is consistent with a gaussian (at 90\% confidence
level). We also note that the predicted number of X-ray
afterglows with a flux lower than about $2 \times 10^{-13}$ is
3.5, consistent with the observed number of objects.}}
\end{figure}

\end{document}